\documentclass[twoside]{dis08}
\usepackage[latin1]{inputenc}
\usepackage[dvips]{graphicx,epsfig,color}
\usepackage{wrapfig,rotating}
\usepackage{amssymb,amsmath,array}

\usepackage{amsmath}
\usepackage{amssymb}
\usepackage{url}
\usepackage{hyperref}
\usepackage{xspace}
\usepackage{graphicx}

\usepackage{a4wide}

\newcommand{\be}{\begin{equation}}
\newcommand{\ee}{\end{equation}}
\newcommand{\bea}{\begin{eqnarray}}
\newcommand{\eea}{\end{eqnarray}}
\newcommand{\bi}{\begin{itemize}}
\newcommand{\ei}{\end{itemize}}
\newcommand{\ben}{\begin{enumerate}}
\newcommand{\een}{\end{enumerate}}

\newcommand{\lp}{\left(}
\newcommand{\rp}{\right)}
\newcommand{\aq}{\alpha_s\left( Q^2 \right)}

\newcommand{\dy}{\ttt{dy}}
\newcommand{\dlnlnQ}{\ttt{dlnlnQ}}

\newcommand{\GeV}{\;\mathrm{GeV}}
\newcommand{\TeV}{\;\mathrm{TeV}}

\newcommand{\hoppet}{\textsc{hoppet}\xspace}
\newcommand{\ttt}[1]{\texttt{#1}}
\newcommand{\order}[1]{{\cal O}\left(#1\right)}
\newcommand{\fn}{\scriptsize}

\usepackage{color}
\definecolor{comment}{rgb}{0,0.3,0}
\definecolor{identifier}{rgb}{0.0,0,0.3}

\usepackage{listings}
\begin{document}
\title{The HOPPET NNLO parton evolution package}

%
%
%
%

\author{Gavin Salam and \underline{Juan Rojo}
\thanks{This work was supported  by grant ANR-05-JCJC-0046-01 from the
French Agence Nationale de la Recherche. }
%
\vspace{.3cm}\\
%
Laboratoire de Physique Theorique et Hautes Energies (L.P.T.H.E.) \\
UPMC Paris VI and Universit\'e Paris Diderot Paris VII \\
 4 place Jussieu, F-75252 Paris Cedex 05 France 
}

\maketitle

\begin{abstract}
 This contribution describes the HOPPET
 Fortran~95 package for carrying out DGLAP
  evolution and other common manipulations of PDFs. 
 The PDFs are represented on a grid in $x$-space so
  as to avoid limitations on the functional form of input
  distributions.  Good speed and accuracy are obtained through the
  representation of splitting functions in terms of their convolution
  with a set of piecewise polynomial basis functions.
We briefly describe the structure of the code and discuss its
quantitative performance. Finally we  comment of future
directions for the program's development.
\end{abstract}

\paragraph{Introduction}
There has been considerable discussion over the past years
of numerical solutions of the 
DGLAP
 equation  
for the QCD
evolution of parton distributions (PDFs),
\begin{equation}
  \label{eq:dglap}
  \frac{\partial q_i(x,Q^2)}{\partial \ln Q^2} = \frac{\aq}{2\pi}
 \int_x^1 \frac{dz}{z}
  P_{ij}\lp z,\aq\rp q_j\lp \frac{x}{z},Q^2\rp\ .
\end{equation}
There exist two main classes of approaches: those that
solve the equation directly in $x$-space and those that solve it for
Mellin transforms of the
parton densities, defined as
\be
\label{eq:mellin}
q_{N}\lp N,Q^2\rp = \int_0^1 dx x^N q_i(x,Q^2) \ ,
\ee 
 and subsequently invert the transform back to
$x$-space.
$N-$space based methods are of interest because the Mellin transform
converts the convolution of eq.~(\ref{eq:dglap}) into a multiplication,
resulting in a continuum of independent matrix differential (rather
than integro-differential) equations, one for each value of $N$,
making the evolution more efficient numerically.

The drawback of the Mellin method is that one needs to know the Mellin
transforms of both the splitting functions and the initial conditions.
The $x$-space method is in contrast more flexible, since the inputs
are only required in $x$-space; however it is generally considered 
less efficient numerically, because of the need to carry out the
convolution in eq.~(\ref{eq:dglap}).

Despite it being more difficult to obtain high accuracy with $x$-space
methods, their greater flexibility means that they are widespread,
serving as the basis of programs like QCDNUM~\cite{Botje}
or CANDIA~\cite{Candia},
and  used also by the CTEQ and MSTW
global fitting
collaborations. A novel approach  which combines
advantages of the $N-$space and $x-$space methods 
is discussed in \cite{nnpdf}.

\hoppet \cite{hoppet}, 
the program described in this contribution \cite{url}, 
uses higher-order methods both for
the $x$-integrations and $Q$ evolution. Apart
from its mainstream PDF evolution
capabilities, it also provides
access to a range of low and medium-level operations on PDFs which
should allow a user to extend the facilities already provided.

The functionality of \hoppet has been present in
its predecessors for several years, 
and a first public release with complete
documentation was recently presented \cite{hoppet}.
Those predecessors have  been used in a number of different
contexts, like resummation of event shapes
in DIS \cite{DisResum}, automated resummation
of event shapes \cite{caesar}, studies of
resummation in the small-$x$ limit
and in a posteriori inclusion
of PDFs in NLO final-state calculations \cite{Banfi:2007gu}, as well
as  used for benchmark
comparisons with Pegasus~\cite{Pegasus} in \cite{Benchmarks}.

\paragraph{Program structure}

The numerical techniques upon which \hoppet is based are presented
in \cite{hoppet}. Let us only recall here that it
uses higher-order methods both for
the $x$-integrations and $Q$ evolution. It combines this with multiple
grids in $x$-space: a high-density grid at large $x$ where it is
hardest to obtain good accuracy, and coarser grids at smaller $x$
where the smoothness of the PDFs facilitates the integrations. One of
the other crucial features of the program is that it pre-calculates as
much information as possible, so as to reduce the 
evolution of a new PDF initial condition to a modest set of addition
and multiplication operations.

The functionalities in \hoppet can be accessed from two interfaces. The first 
one is a general interface, available in F95 only,
 required for
those who want to delve into the innards of the program, and which provides
access  to a range of
low and medium-level operations allow  to extend 
\hoppet basic capabilities.  The essential \hoppet 
functionalities are also accessible from a streamlined interface,
which can be accessed from F97, F95 and C++.

\begin{table}
\begin{center}
{\bf \large STREAMLINED INTERFACE}
\\
\vspace{0.2cm}
\begin{tabular}{|c|c|}
\hline
\bf METHOD  & \bf DESCRIPTION \\
\hline
\bf Initialisation & \\
\hline
\begin{lstlisting}
 hoppetStart(dy,nloop)
\end{lstlisting} 
& \scriptsize
Sets up a compound grid with
spacing in $\ln 1/x$ of \ttt{dy} at small $x$,\\ &
\fn~  extending to $y = 12$ and numerical
order $\ttt=-5$. \\ & \fn~  The $Q$ range for the tabulation will be $1\GeV <
Q<28 \TeV$, \\
& \fn~ \ttt{dlnlnQ=dy/4} and the factorisation scheme is ${\overline{\rm MS}}$\\
\hline
\begin{lstlisting}
 hoppetStartExtended(ymax,dy,Qmin,
 Qmax,dlnlnQ,nloop,order,factscheme)
\end{lstlisting} & \fn
  ~More general initialisation \\
\hline 
\begin{lstlisting}
 hoppetSetFFN(fixed_nf)
 hoppetSetVFN(mc, mb, mt)
\end{lstlisting} &
\fn ~Set heavy flavour scheme
\\
\hline
\begin{lstlisting}
  alphas = hoppetAlphaS(Q)
\end{lstlisting} &
\fn~Accessing the coupling \\
\hline&\\[-0.5em]
\bf Normal evolution & \\
\hline
\texttt{\footnotesize
 hoppetEvolve(asQ,Q0alphas,}
& \fn\ PDF evolution: specifies the coupling \ttt{asQ} at a 
scale \ttt{Q0alphas}, \\
 \texttt{\footnotesize nloop,muR\_Q,LHAsub,Q0pdf)}&
 \fn the number of loops for  evol., \ttt{nloop}, \\
&\fn the ratio (\ttt{muR\_Q}) of ren. to fact. scales. \\ 
&  \fn the name of a subroutine \ttt{LHAsub} with an LHAPDF-like interface \\
&  \fn and the scale
\ttt{Q0pdf} at which one starts the PDF evolution \\
&  \fn Note:  \ttt{LHAsub} only called at scale \ttt{Q0pdf}\\
\hline
\begin{lstlisting}
 hoppetEval(x,Q,f)
\end{lstlisting} &
\fn On return, \ttt{f(-6:6)} contains all flavours of the PDF set\\ 
& \fn ~
(multiplied by $x$) at the given $\ttt{x}$ and $\ttt{Q}$ values \\
\hline&\\[-0.5em]
\bf Cached evolution & \\
\hline
\begin{lstlisting}
 hoppetPreEvolve(asQ,Q0alphas, 
          nloop, muR_Q, Q0pdf)
\end{lstlisting} & \fn ~
 Preparation of the cached evolution\\
\hline
\begin{lstlisting}
 hoppetCachedEvolve(LHAsub)
\end{lstlisting} &
\fn  Perform cached evolution with the initial condition\\
& \fn at \ttt{Q0pdf} from a routine \ttt{LHAsub} 
with LHAPDF-like interface\\
&  \fn Notice  \ttt{LHAsub} only called at scale \ttt{Q0pdf}\\
\hline
\begin{lstlisting}
 hoppetEval(x,Q,f)
\end{lstlisting} &
\fn On return, \ttt{f(-6:6)} contains all flavours of the PDF set\\ 
& \fn ~
(multiplied by $x$) at the given $\ttt{x}$ and $\ttt{Q}$ values \\
& \fn [as for normal evolution]\\
\hline
\end{tabular}
\end{center}
\caption{\label{tab:streamlined} Reference guide for the streamlined
  interface.}
\end{table}

A summary of the most important modules in the streamlined interface
can be seen in Table \ref{tab:streamlined}. It is clear that, thanks to the
structure of this interface, the evolution of PDFs with a large
variety of options can be performed with few selected routines.
The analog reference guide for the general interface can be found
in \cite{hoppet}.

 Often one simply wishes to provide  PDFs
at some initial scale and then subsequently be able to access it at
arbitrary values of $x$ and $Q$. For this purpose it is useful 
(and most efficient) in \hoppet to
produce a table of the PDFs as a function of $Q^2$, which then allows
for access to the PDFs at arbitrary $x, Q$ using an interpolation.

It is also possible to prepare an evolution in cached form. This is
useful if one needs to evolve many different PDF sets with the same
evolution properties (coupling, initial scale, etc.), as is the
usual situation in global analyses of PDFs, 
the initialisation may take a bit longer than a normal evolution
($2$--$10$ times depending on the perturbative order), 
however, once it is done,
cached evolutions run $3$--$4$ faster than a normal evolution.

\paragraph{Performance}

 \hoppet's
correctness has been established with a reasonable degree of
confidence in the benchmark tests~\cite{Benchmarks} where it was
compared with the Mellin space based
evolution code QCD-Pegasus \cite{Pegasus}.

For use in most physical applications, an accuracy in the
range $10^{-3}$ to $10^{-4}$ is generally  more than adequate. 
The critical issue in such cases is more likely to
be the speed of the code, for example in PDF fitting
applications.
\hoppet's accuracy and speed both depend on the choice of grid (in
$y=\ln 1/x$) and the evolution and/or tabulation steps in $Q$.

\begin{figure}
  \centering
 \includegraphics[width=0.48\textwidth]{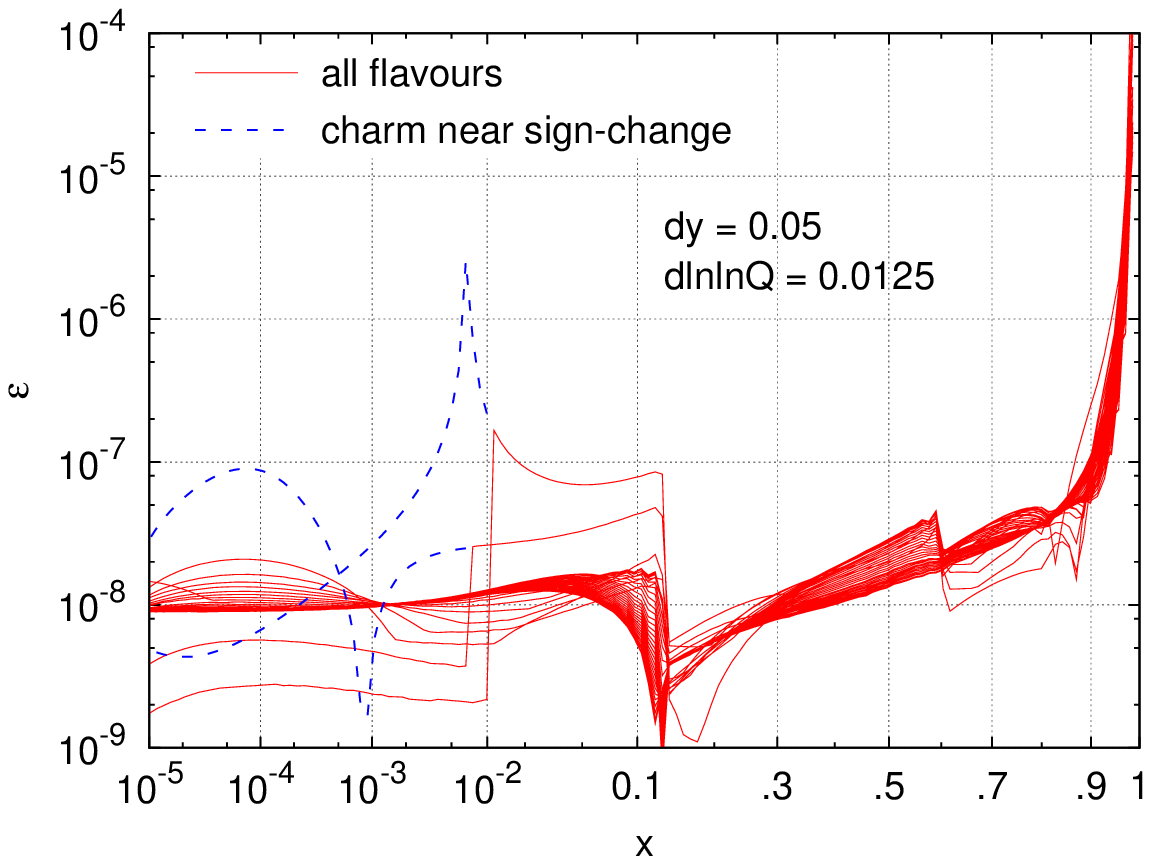}
  \includegraphics[width=0.48\textwidth]{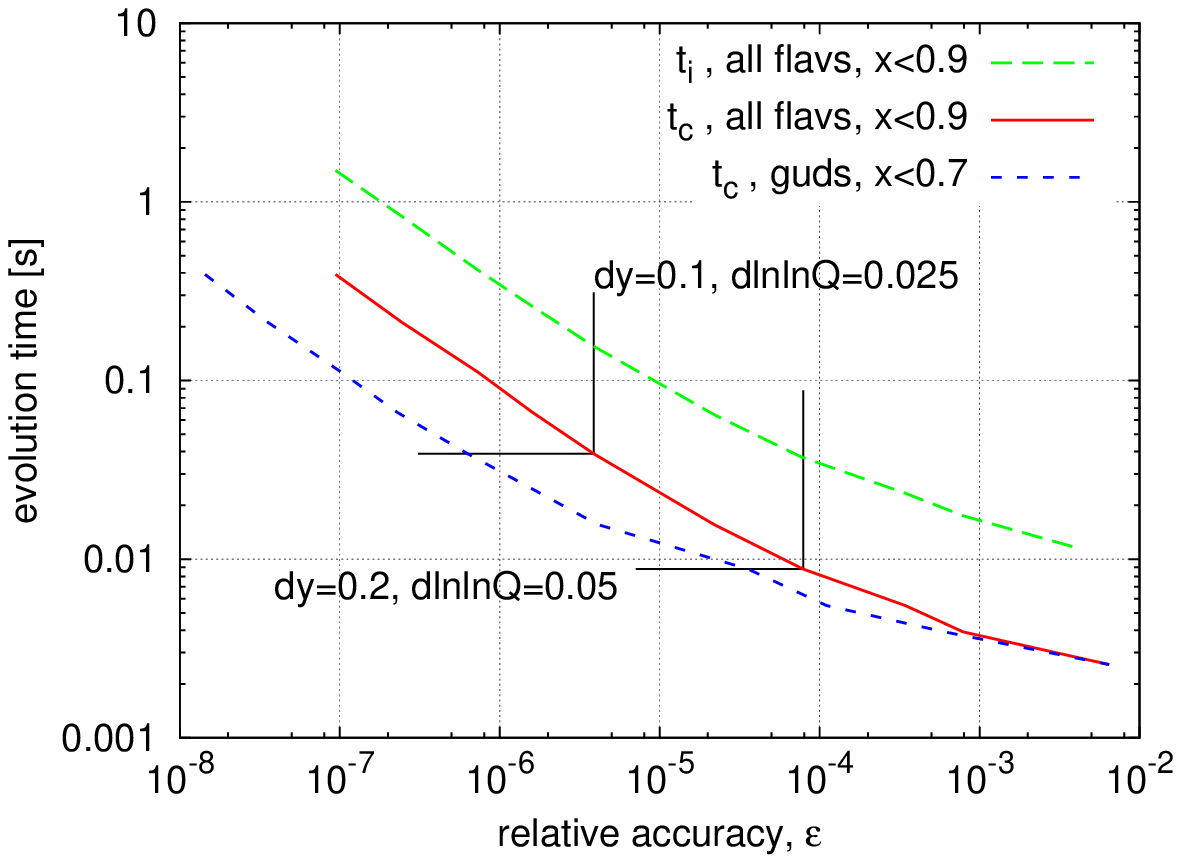}
  \caption{Left: the relative accuracy $\epsilon$ of the least well
    determined flavour channel at each $x, Q$ point, shown as a
    function of $x$ for many $Q$ values. 
Right: the same quantity obtained as a function of the time taken
    to perform the evolution.}
  \label{fig:acc-v-time}
\end{figure}

Fig.~\ref{fig:acc-v-time} shows the relative accuracy
$\epsilon$ as a function of $x$ for $\dy=0.05$ and $\dlnlnQ=\dy/4$.
The relative accuracy $\epsilon$
is poorest as
one approaches $x=1$, where the PDFs all go to zero very rapidly and
so have divergent logarithmic derivatives in $x$,
adversely affecting the accuracy of the convolutions. This region is
always the most difficult in $x$-space methods, however the use of
multiple subgrids in $x$ allows to one to obtain good results
for $x<0.9$ which is likely to be the largest value of any
phenomenological relevance.

The time spent in \hoppet for a given analysis can expressed as
follows, according to whether or not one carries out cached
pre-evolution:
\begin{subequations}
  \label{eq:timing}
  \begin{align}
    t_\text{no pre-ev}   &= t_s + n_\alpha t_\alpha + n_i (t_i  + n_{xQ}\, t_{xQ})\,,\\
    t_\text{with pre-ev} &= t_s + n_\alpha (t_\alpha + t_p) + n_i (t_c + n_{xQ}\,
    t_{xQ})\,,
  \end{align}
\end{subequations}
where $t_s$ is the time for setting up the splitting functions,
$n_\alpha$ is the number of different running couplings that one has,
$t_\alpha$ is the time for initialising the coupling,
$n_i$ is the number of PDF initial conditions that one wishes to
consider, $t_i$ is the time to carry out the tabulation for a single
initial condition, $n_{xQ}$ is the number of points in $x,Q$ at which
one evaluates the full set of flavours once per PDF initial condition;
in the case with pre-prepared cached evolution, $t_p$ is the time for a
preparing a cached evolution and $t_c$ is the time for performing the
cached evolution. Finally $t_{xQ}$ is the time it takes to evaluate
the PDFs at a given value of $(x,Q^2)$ once the tabulation has
been performed.

\begin{table}
  \centering
  \begin{tabular}{|ll|c|c|c|}\hline
          &&     lf95  &  ifort   & g95    \\\hline
   $t_s$ &[s]  &  0.9  &  0.66    & 2.8    \\
   $t_\alpha$ & [ms]                
               &  0.16 &  0.12    & 0.13   \\
   $t_i$ &[ms] &  37   &   38     & 330    \\
   $t_p$ &[ms] &  51   &   44     & 310    \\
   $t_c$ &[ms] &  8.8  &   9.8    & 110    \\
   $t_{xQ}$ &[$\mu$s]           
               &  2.7  &   3.1    &  25    \\
   \hline
  \end{tabular}
  \caption{Contributions to the run time in eqs.~(\ref{eq:timing})
 for    $\dy=0.2$ and 
    $\dlnlnQ=0.05$ and standard values for the other parameters 
    (on a 3.4GHz Pentium IV (D) with 2~MB cache).}
  \label{tab:timings}
\end{table}

The various contributions to the run-time are shown in
Table~\ref{tab:timings} for $\dy=0.2$ and $\dlnlnQ=0.05$ (giving an
accuracy $\sim 10^{-4}$), for various compilers.  In a typical
analysis where run-times matter, such as a PDF fit, it is to be
expected that the time will be dominated by $t_c$ (or $t_i$). However,
in the typical case of global PDF fits, for which
the number of $x,Q$ points is rather large ($\gtrsim 3000$), 
 it will be $n_{xQ} t_{xQ}$ that takes the most time\footnote{
With these numbers, it is easy to check that a global fit with
$n_{xQ}\sim 3000$ and $n_i\sim 10^5$ could be completed
in less than half an hour.}.

We study $t_c$ and $t_i$ in more detail in Fig.~\ref{fig:acc-v-time},
where we relate them to the accuracy obtained from the evolution. As
one would expect, studying just the `guds' flavours for $x<0.7$ one
obtains better accuracy for a given speed than with all flavours for
$x<0.9$. Overall one can obtain $10^{-4}$ accuracy with $t_c \simeq
10^{-2}$\,s and $10^{-6}$ accuracy with $t_c \simeq 10^{-1}$\,s.

The timings shown here are roughly similar, for accuracies $\sim
10^{-4}$, to those obtained with the $N$-space code
Pegasus~\cite{Pegasus} when the number of $x,Q$ points to be
evaluated is $\order{10^3}$.
To the best of our knowledge,
 other NNLO evolution codes published in recent years 
 seem generally less competitive
either in terms of accuracy or speed.

\paragraph{Future directions}

\hoppet is an $x$-space evolution code that is novel both in terms of
the accuracy and speed that it provides compared to other $x$-space
codes, and in terms of its interface, designed to provide a
straightforward and physical way of manipulating PDFs beyond the
built-in task of DGLAP evolution.

Work in progress for future releases include the
computation of deep-inelastic scattering
structure functions and reduced cross sections.
Furthermore, a physical feature absent from mainstream PDF
evolution codes is that of evolution that includes matching with
various types of resummed calculations. 
There is work in progress to implement the ABF and CCSS
small-$x$ resummations \cite{Altarelli:2008aj}
into \hoppet through the ability to
manipulate  general, user-defined, interpolated splitting
functions and coefficient functions.

\begin{footnotesize}

\end{footnotesize}


\end{document}